\documentstyle[equations,12pt]{article}
\newcommand{\ep}{\epsilon}
\newcommand{\lab}{\label}
\newcommand{\non}{\nonumber}

\newcommand{\be}{\begin{equation}}
\newcommand{\ee}{\end{equation}}
\newcommand{\bea}{\begin{eqnarray}}
\newcommand{\eea}{\end{eqnarray}}
\def\Journal#1#2#3#4{{#1} {\bf #2}, {#3} {(#4)}}

\def\JMP{\em J. Math. Phys.}

\def\PLB{{\em Phys. Lett.}  \bf B}
\def\PRD{{\em Phys. Rev.}  \bf D}

\def\lab{\label}
\begin{document}
\setcounter{page}{0}
%
%
%
{\Large\bf The symmetry structure of the heavenly equation}

\vspace{.3cm}
{\large E. Alfinito \footnote{alfinito@le.infn.it}, G. Soliani 
\footnote{soliani@le.infn.it} and L. Solombrino \footnote{solombrino@le.
infn.it}\\
{\it Dipartimento di Fisica dell'Universit\`a, 73100 Lecce,
Italy,}\\ {\it and Istituto Nazionale di Fisica Nucleare, Sezione di Lecce, 
Italy.}}

\vspace{.3cm}

\date{April 1996}

{\small
\begin{abstract}
We show that excitations of physical interest of the heavenly equation 
are generated by 
symmetry operators which yields two reduced equations with different 
characteristics. One equation is of the Liouville type and gives rise to 
gravitational instantons, including those found by Eguchi-Hanson and 
Gibbons-Hawking. The second equation appears for the first time in the 
theory of heavenly spaces and provides meron-like configurations endowed 
with a fractional topological charge. 
A link is also established between the heavenly 
equation and the socalled Schr{\"o}der equation, which plays a crucial role 
in the bootstrap model and in the renormalization theory.
\end{abstract}}

\vspace{.3cm}


\vspace{.3cm}

An important statement concerning self-dual, euclidean, Einstein spaces 
with one rotational Killing vector (heavens), is that these are completely 
determined by a real scalar field obeying a nonlinear partial differential 
equation which formally coincides with a continuous version of the Toda 
lattice \cite{Saveliev}. This equation can be written as
\be u_{xx}\,+\, u_{yy}\,=\,\kappa(e^{u})_{zz}
\lab{1} \ee
where the signature factor $\kappa\,=\,\pm1$ has been introduced for later 
convenience.

Equation (\ref{1}) appears in a variety of physical areas, ranging from the 
theory of Hamiltonian systems to general relativity. In the last 
context, the metric tensor related to Eq. (\ref{1}) is
\bea g\,&=&\,{u}_{z}\left\{e^{u}[(dx)^{2}\,+\,(dy)^{2}]\,+\,(dz)^{2}
\right\}\,\\ \non
&+&\,
\frac{1}{u_{z}}\left[\epsilon( -u_{y}dx\,+\,u_{x}dy)\,+\,d\tau
\right]^{2},
\lab{2} \eea
where $\ep\,=\,\pm 1$ for self-dual and anti-self-dual cases, respectively.

Heavenly spaces have been studied by several authors in different 
frameworks \cite{Plebanski}. The main aspects of these investigations concern
{\it i)} the problem of the classification of the Riemann metrics admitting 
at least a Killing vector field; {\it ii)} the discover of extended 
conformal symmetries, i.e. the large $n$ limit of the Zamolodchikov $ W_{n}$ 
algebra \cite{Saveliev};
{\it iii)} the search of exact solutions of 
physical significance, such as self-dual gravitational instantons 
\cite{Eguchi}, whose role is important in quantum gravity \cite{Hawking}.

In this note, we focus the attention on both the symmetry structure and the 
determination of new exact solutions of Eq.(1). In regard to the last point, 
even particular solutions are hard to find. An effective method to shed 
light on the symmetry properties of Eq.(1) and, correspondingly, to 
construct explicit configurations is based on the reduction approach 
\cite{Olver}, which exploits group-theoretical techniques. Following these 
ideas, here we show that Eq.(1) allows an infinite-dimensional Lie algebra, 
a realization of which is given by generators whose coefficients are 
harmonic functions in the variables $x,y$. Some generators, say $X_{n}$, 
obey the Virasoro algebra $\cal L$
without central charge. The remaining generators, 
say ${\tilde X}_{n}$, which are constructed by the harmonic functions 
conjugate to those related to the Virasoro operators, satisfy a set of 
commutation relations involving the presence of both $X_{n}$ and 
${\tilde X}_{n}$.

Furthermore, we find two reduced equations which furnish two classes of 
solutions to Eq.(1) with quite different properties. One of them, 
Eq.(\ref{11}), provides meron-like gravitational configurations which allow a 
fractional topological charge. The second equation, (\ref{15-b}), yields 
instanton configurations (with integer topological numbers) containing the 
Eguchi-Hanson and the Gibbons-Hawking 
gravitational instantons as particular cases. Finally, an 
unexpected connection is shown between the heavenly equation and the 
socalled Schr{\"o}der equation which is important in the statistical 
bootstrap model and in renormalization theory \cite{Hagedorn}.

It turns out also that after a suitable change of basis, the 
elements $X_{-1},\,X_{0},\,X_{1},\,{\tilde X}_{-1},\,{\tilde X}_{0},\,
{\tilde X}_{1}$ of $\cal L$ 
fulfil the commutation relations of the algebra of the proper 
Lorentz group $SO(3,1)$, which is locally isomorphic to $SL(2,C)$ 
\cite{Carmeli2}. Thus, 
the appearance of $SL(2,C)$ in the theory of the heavenly spaces is a 
direct consequence of the symmetry structure of Eq.(1). This property is 
also partaken by the Newman-Penrose equations, i.e. the null tetrad 
formulation of the Einstein field equations which are invariants under the 
group $SL(2,C)$. Therefore, the choice of $SL(2,C)$ as the gauge group used 
in the local gauge approach to Einstein's general theory of relativity 
\cite{Carmeli} arises in a natural way by symmetry considerations.

The symmetry algebra of Eq.(1) can be obtained applying the procedure of 
Ref.\cite{Olver}. The generator of the corresponding symmetry group has 
been found in part with the help of a computer, by using the symbolic 
language REDUCE \cite{SC}. Our result is that Eq.(1) admits an infinite 
dimensional Lie group of symmetries, that is a local group $G$ of 
transformations acting on the independent variables $(x,y,z)$ and the 
dependent variable $u$, with the property that whenever $u(x,y,z)$ is a 
solution of Eq.(1), then $u^{\prime}\,=\,(g\circ u)(x^{\prime},y^{\prime},
z^{\prime})$ is also a solution for any $g\,\in G$. The symmetry algebra 
underlying $G$ is represented by the vector fields
\begin{subequations} \label{5}
       \begin{eqalignno}
X(f)&= f\,\partial_{x}\,+\,{\hat f}\,\partial_{y}\,-\,2f_{x}\,\partial_{u}
, \label{5-a}\\
Y & = z\partial_{z}\,+\,2\partial_{u}, \label{5-b}\\
T&= \partial_{y}, \label{5-c}\\
Z&= \partial_{z}, \label{5-d}
\end{eqalignno}
\end{subequations}
where $f\,=\,f(x,y)$ is an arbitrary harmonic function, and ${\hat f}\,=\,
{\hat f}(x,y)$ is its conjugate.

We observe that the translation operator along the axis $x$ is contained in 
(\ref{5-a}), in the sense that $\partial_{x}\,\equiv\,X(1)$. The generators 
(\ref{5}) obey the commutation relations
\begin{subequations} \label{6}
       \begin{eqalignno}
{[X(f),X(g)]}= X(fg_{x}-gf_{x}+{\hat g}{\hat f}_{x}-{\hat f}
{\hat g}_{x}), \label{6-a}\\
{[X(f),T]} = - X(f_{y}),\;\; 
{[Z,Y]}= Z,\qquad\label{6-c}\\
{[T,Z]}\,=\,{[X(f),Z]}\,
=\,{[T,Y]}\,  =\,[X(f),Y] \,=\,0. \label{6-d}
\end{eqalignno}
\end{subequations}
The presence of (harmonic) arbitrary functions tells us that the symmetry 
algebra (\ref{6}) is infinite-dimensional. This algebra can be embedded into 
a loop algebra of the Virasoro type without central charge. This can be 
shown by assuming that $f$ and $g$ are of the form $f_{n}\,=\,-e^{nx}\,
\cos ny$ and $g_{m}\,=\,-e^{mx}\,\cos my$ ($n,m\;\in {\cal Z}$). 
In such a way, we 
build up a sub-algebra of the symmetry algebra (\ref{6}) with the basis $Y,T,
Z$, and
\begin{subequations} \label{6.2}
       \begin{eqalignno}
X_{n}=X(f_{n})=e^{nx}\left(-\cos{ny}\partial_{x}
 -\sin{ny}\partial_{y}+2n \cos{ny} \partial_{u}\right),\label{6.2-a}\\
{\tilde X}_{n}= X({\hat f}_{n})\,=\,e^{nx}\left(-\sin{ny}\partial_{x}
+\cos ny\partial_{y} +2n \sin ny \partial_{u}\right).
\label{6.2-b}
\end{eqalignno}
\end{subequations}
The commutation relations
\begin{subequations} \label{6.3}
       \begin{eqalignno}
{[X_{n},X_{m}]}&=(n-m) X_{n+m}, \quad 
{[X_{n},T]}  = n {\tilde X}_{n},\label{6.3-a}\\ 
{[X_{n},{\tilde X}_{m}]}&= (n-m) {\tilde X}_{n+m}, \label{6.3-c}\\
{[{\tilde X}_{n},{\tilde X}_{m}]}&= - (n-m) { X}_{n+m}, \label{6.3-d}\\
{[{\tilde X}_{n},T]}&= -n {X}_{n},\quad 
{[{\tilde X}_{n},Y]}= [{\tilde X}_{n},Z]\,=\,0, \label{6.3-f}
\end{eqalignno}
\end{subequations}
hold, together with (\ref{6-c}) and (\ref{6-d}).

The existence of the Virasoro-type algebra (\ref{6.3}) reflects the 
invariance property of Eq.(1) under a conformal transformation. In other 
words, if $u(x,y,z)$ is a solution of Eq.(1), the function $\tilde{u}\,=\,
u(x^{\prime},y^{\prime},z)$ given by
\be
\tilde{u} \,=\,u(x,y,z)\,+\,\ln(U^{2}_{x^{\prime}}+V^{2}_{x^{\prime}}),
\lab{6-bis}\ee
is also a solution,
where $x=U(x^{\prime},y^{\prime})$ and $y=V(x^{\prime},y^{\prime})$ are any 
pair of conjugate harmonic functions.

Equation (\ref{6-bis}) is a direct consequence of the invariance of Eq.(1), 
written in the complex form
\be
4\partial_{\eta}\partial_{\bar\eta}u\,=\,\kappa \partial^{2}_{z}e^{u},
\lab{6-ter}\ee
under the transformations
$$ \eta = f(\zeta),\quad \bar{\eta} = \bar{f}(\bar{\zeta}),\quad u = \tilde{u}
- \ln(f^{\prime}\bar{f}^{\prime}),$$
where $\eta = x+ iy$,  $\bar\eta = x- iy$, $\partial_{\eta} =\frac{1}{2}
(\partial_{x}- i\partial_{y})$, $\partial_{\bar\eta} =\frac{1}{2}
(\partial_{x}+ i\partial_{y})$, $f^{\prime}=f_{\zeta}(\zeta)$, 
$\bar{f}^{\prime}=f_{\bar\zeta}(\bar\zeta)$ and $f(\zeta)$ is an arbitrary 
holomorphic function of $\zeta= x^{\prime}+iy^{\prime}$.
  
It is noteworthy that the commutation relations (\ref{6.3}) contain those of 
the proper Lorentz group $SO(3,1)$. This can be seen by putting
$$
J_{1}\,=\,\frac{i}{2}(X_{1}-X_{-1}),\;J_{2}\,=\,-\frac{1}{2}(X_{1}+X_{-1}),\;
$$ 
$$J_{3}\,=\,-{i}X_{0},$$
$$P_{1}\,=\,\frac{i}{2}({\tilde X}_{1}-{\tilde X}_{-1}),\;
P_{2}\,=\,-\frac{1}{2}({\tilde X}_{1}+{\tilde X}_{-1}),\;$$
$$P_{3}\,=\,-{i}{\tilde X}_{0}.$$
Then,
\begin{eqalignno}
{[J_{i},J_{j}]}\,= & \,\ep_{ijk} J_{k},\;
{[P_{i},P_{j}]}\,=\,-\ep_{ijk} J_{k},\; \nonumber\\
{[J_{i},P_{j}]}\,= & \,\ep_{ijk} P_{k}.
\lab{7}
\end{eqalignno}

Some important solutions which can be identified with meron-like 
configurations and
gravitational 
instantons emerge applying the method of symmetry reduction to Eq.(\ref{1})
. This amounts essentially to finding the invariants (symmetry variables) 
of a given subgroup of the symmetry group admitted by Eq.(\ref{1}). A basis 
set of symmetry variables $I(x,y,z,u)$ for each vector field $X$ is 
obtained by solving the first order partial differential equation 
$XI(x,y,z,u)\,=\,0$. Below some cases of special physical interest are 
considered. Let us start from the symmetry generator
\be X(f)\,=\,x\partial_{x}\,+\,y\partial_{y}\,-\,2\partial_{u},
\lab{8}\ee 
$(f\,=\,x,\hat{f}\,=\,y)$. A set of basis invariants is given by
$$ z\,=\,z^{\prime},\qquad t\,=\,\frac{y}{x},$$
\be W(z,t)\,=\,2\ln x\,+\,u(x,y,z). \lab{8-bis} \ee
Then, substitution from (\ref{8-bis}) into Eq.(\ref{1}) yields
\be (1\,+\,t^{2}) W_{tt}\,+\,2t\,W_{t}\,+\,2\,=\,\kappa 
(e^{W})_{zz}.
\lab{9}\ee
Exact solutions to Eq.(\ref{9}) can be provided in the framework of group 
analysis. In doing so, we have that Eq.(\ref{9}) admits the Lie-point 
symmetry vector fields
\begin{subequations} \label{10}
       \begin{eqalignno}
Z_{0}\,&=\,z\partial_{z}\,+\,2\partial_{W},
\label{10-a}\\
T_{0}\,&=\,\partial_{z},\label{10-b}\\
X_{0}\,&=\,(1+t^{2})\,{\rm arctan}t\partial_{t}
-2(1\,+\,t\,
{\rm arctan}t)\partial_{W}, \label{10-c}\\
Y_{0}\,&=(1+t^{2})\,\partial_{t}\,-\,2t\partial_{W},\label{10-d}
\end{eqalignno}
\end{subequations}
$$ {\rm with}\qquad 
[T_{0},Z_{0}]\,=\,T_{0},\;[Y_{0},X_{0}]\,=\,Y_{0},$$
$$[T_{0},X_{0}]\,=\,
[T_{0},Y_{0}]\,=\,[Z_{0},X_{0}]\,=\,[Z_{0},Y_{0}]\,=\,0.$$

The two-dimensional equation (\ref{9}) can be reduced to an ordinary 
differential equation in correspondence of any linear combination of the 
symmetry variables (\ref{10}). For example, the operator $Z_{0}$ generates 
a configuration which can be interpreted as a meron-like excitation.
In fact, the reduced equation related to $Z_{0}$ is
\be (1+t^{2})\sigma_{tt}\,+\,2t\sigma_{t}\,=\,1\,-\,\kappa e^{-2\sigma}
,\lab{11}\ee
where $\sigma\,=\,\sigma(t)\,=\,\ln z\,-\,\frac{1}{2}W(z,t)$, 
$z,\;t\,=\,\frac{y}{z}$ are invariants. 

Equation (\ref{11}) can be exactly solved by introducing a function 
$\psi(t)$ such that
\be
\psi_{t}\,=\,1-\kappa e^{-2\sigma}.
\lab{11a}\ee
Then, Eq.(\ref{11}) can be integrated to give
\be
(1+t^{2})\psi_{tt}\,=\,2(c+\psi)(1-\psi_{t}),
\lab{11aa}\ee
where $c$ is an arbitrary constant.

Now, by setting $\phi=t-c-\psi$, Eq.(\ref{11aa}) becomes the Riccati equation
\be
(1+t^{2})\phi_{t}\,=\,\phi^{2}\,+\,c_{0}^{2},
\lab{11b}\ee
($c_{0}$= constant) which yields the equation of the hypergeometric type 
\cite{Gatteschi}
\be
\gamma_{tt}\,+\,\frac{2t}{1+t^{2}}\,\gamma_{t}\,+\,
\frac{c^{2}_{0}}{(1+t^{2})^{2}}\,\gamma\,=\,0,
\lab{11c}\ee
via the transformation
\be
\phi\,=\,-(1+t^{2})\frac{\gamma_{t}}{\gamma}.
\lab{11d}\ee
The general solution of (\ref{11c}) is
\be
\gamma\,=\,A\left(\frac{1-it}{1+it}\right)^{a}\,+\,
B\left(\frac{1+it}{1-it}\right)^{a},
\lab{11e}\ee
where $a\,=\,c_{0}/{2}$ and $A,B$ are two constants of integration. 

With the help of (\ref{11a}), (\ref{11d}) and (\ref{11c}) we have
         \be
e^{-2\sigma}\,=\,\kappa \phi_{t}=\kappa\left[\frac{4a^{2}}{1+t^{2}}\,
+\,(1+t^{2})\left(\frac{\gamma_{t}}{\gamma}
\right)^{2}
\right],
            \lab{11f}\ee
which can be cast into the explicit form
           \be
e^{-2\sigma}\,=\,16 \kappa a^{2}\cos^{2}\theta 
\frac{AB}{ \left( A e^{-2ia\theta} + B e^{2ia\theta}\right)^{2} },
\lab{11g}\ee
in polar coordinates $x=r \cos \theta$, $y=r \sin \theta$.

The class of real $\sigma$ corresponds to $A=B^{*}$ and $a^{2}>0$ for 
$\kappa=1$, and to $A=B^{*}$ and $a^{2}<0$ for $\kappa=-1$.
Some particular cases are of special interest. For example, for 
$a=\frac{1}{4}$ and $A=B^{*}=1+i$, from (\ref{11g}) we find 
($\kappa=1$)
\be
e^{-2\sigma}\,=\,\frac{1}{2}\left(1 \pm \frac{t}{\sqrt{1+t^{2}}}
\right).
\lab{11h}\ee
The presence of the sign $\pm$ is due to the invariance of Eq.(\ref{11}) 
under the change of variable $t\rightarrow -t$. Consequently, the 
transformation $(\ref{8-bis})$ provides
\begin{subequations} \label{13}
       \begin{eqalignno}
u_{\pm}\,=\,\ln \frac{z^{2}}{2x^{2}}\left(1\,\pm\,
\frac{y}{\sqrt{x^{2}+y^{2}}}
\right).
\label{13-a}
\end{eqalignno}
This expression can be elaborated by resorting to the invariance property 
of Eq.(1) under the conformal transformation (\ref{6-bis}). Indeed, by setting 
in (\ref{13-a}) $x\rightarrow 2xy$, $y\rightarrow x^{2}- y^{2}$ (i.e., by 
passing to a set of parabolic coordinates), we obtain
       \begin{eqalignno}
\tilde{u}_{+}\,=\,\ln \frac{z^{2}}{y^{2}},\lab{13-aa}\\
\tilde{u}_{-}\,=\,\ln \frac{z^{2}}{x^{2}}.
\label{13-b}
\end{eqalignno}
Furthermore, performing in $\tilde{u}_{-}$ the change
to parabolic variables, we are led to the solution
 \begin{eqalignno}
{u}_{0}\,=\,\ln z^{2}\left(\frac{1}{x^{2}}\,+\,\frac{1}{y^{2}}
\right).
\label{13-c}
\end{eqalignno}
\end{subequations}

Now some comments are in order. First, the solution (\ref{13-b}) 
corresponds to the simplest nontrivial solutions of Eq.(\ref{9}) and 
(\ref{11}), namely $e^{W}=z^{2}$, $\sigma=0$ (see (\ref{8-bis})). 
The solution (\ref{13-aa}) corresponds to $e^{W}=z^{2}/t^{2}$, 
$\sigma=\ln t$, while the configuration (\ref{13-c}) corresponds to
\begin{subequations}
\begin{eqalignno}\lab{24}
e^{W}=
z^{2}\frac{1+t^{2}}{t^{2}},\quad &\sigma=\ln\frac{t}{\sqrt{1+t^{2}}}.
\lab{24-a}
\end{eqalignno}
\end{subequations}
Of course, all the abovementioned expressions for $\sigma$ arise from 
(\ref{11g}) as particular cases. We notice that the related 
configurations for $u(x,y,z)$ can be determined via the conformal 
invariance property of Eq.(1) starting from the simplest solution 
$\sigma=0$. Moreover, by comparing (\ref{13-c}) with (\ref{13-aa}) 
and (\ref{13-b}) we have 
$e^{{\tilde u}_{+}}+e^{{\tilde u}_{-}}=e^{u_{0}}$, which looks as a 
"superposition" formula in the variable $e^{u}$.

Second, Eq.(\ref{11}) plays a role similar to that shown by certain {\em meron} 
equations in the context of Yang-Mills theories \cite{Gidas}. This can be 
seen in a heuristic way as follows. For example, let us deal with the 
solution $\sigma=\ln t$, and associate with it the topological charge 
defined by \cite{Cheng} 
        \be
q\,=\,\frac{-i}{2\pi}\int^{\infty}_{-\infty}\,\frac{g^{\prime}(t)}{g(t)} \,
dt,
\lab{25}\ee
where $g^{\prime}=\frac{dg}{dt}$, and $g=e^{\sigma}$. Since $g^{\prime}=
\sigma^{\prime}e^{\sigma}$, the integral (\ref{25}) is divergent. However, 
it is meaningful as Cauchy principal value, where $1/t$ is taken as a 
distribution \cite{Friedlander}. In this sense, $q$ turns out to be zero. 
Anyway, by limiting ourselves to consider the integral of $g^{\prime}(t)/g(t)$ 
over the interval $(-\epsilon, \epsilon)$ $(\epsilon >0)$, we obtain $q=
\frac{1}{2}$. This value of $q$ is characteristic of meron-like 
configurations. On the basis of these considerations, we argue that the 
solution $\tilde{u}_{+}=\ln \frac{z^{2}}{y^{2}}$,
corresponding to $\sigma=\ln t$, behaves as a meron-like excitation rather 
than an instanton-like solution. Furthermore, since the solutions
 (\ref{13-b}) 
and (\ref{13-c}) are amenable to $\tilde{u}_{+}$ via a suitable conformal 
transformation, $\tilde{u}_{-}$ and $u_{0}$ can be interpreted as meron-like 
solutions as well. (Really, through the transformations $x\rightarrow y$, 
$y\rightarrow x$, and
 $x\rightarrow 2xy$, 
$y\rightarrow x^{2}-y^{2}$, from (\ref{13-aa}) 
we obtain (\ref{13-b}) and (\ref{13-c}), 
respectively).

Another feature confirming the physical character of the configurations 
(\ref{13-aa}), (\ref{13-b}) and (\ref{13-c}) is their invariance under the 
transformation $x\rightarrow x/(x^{2}+y^{2})$, 
$y\rightarrow -y/(x^{2}+y^{2})$, which corresponds 
to the inversion $\eta \rightarrow 1/\eta$. We remind 
that, as one expects, this property is shared also by the gravitational 
instantons (\ref{16-b}), (\ref{17-a}) and (\ref{17-b}). Finally, we observe 
that the singularities at $x=0$, $y=0$ of the solutions (\ref{13-aa}), 
(\ref{13-b}) and (\ref{13-c}) can be shifted through the transformations 
$x\rightarrow x+x_{0}$, $y\rightarrow y+y_{0}$ where $x_{0},\;y_{0}$ are 
arbitrary constants (see (\ref{6-bis})).

Now, it is instructive to deal with the generators $J_{1},J_{2}, P_{1},
P_{2}$ of the group $SL(2,C)$. 
%

In correspondence of these operators we obtain four reduced equations of 
the meron-like type, which either coincide formally with (\ref{9}), 
or are led to (\ref{9}) via 
a simple change of the independent variable. Hence, 
these equations bring to eight 
solutions to Eq.(1) which present a meron-like character. For brevity, 
here we shall report only the result for $J_{1}$. By using the set of 
invariants $z, \;\;\xi=\sin{y}/\sinh{x}$, 
$W(z,\xi)\,=\,u\,+\,2 \ln{\sinh{x}},$ we find the reduced equation 
$(1+\xi^{2}) W_{\xi\xi}+2\xi W_{\xi}+2=\kappa (e^{W})_{zz}$, which yields 
the solutions $(\kappa =1)$
\begin{subequations} \label{15}
\begin{eqalignno}
e^{u^{\pm}_{1}}\,=\,
 \frac{z^{2}}{2\sinh^{2}x} \left[1\,\pm\,
\frac{\sin y}{\sqrt{\sinh^{2} x+\sin^{2} y}}
\right].
\label{15-a}
\end{eqalignno}
\par Another symmetry generator important by a physical point of view is $Y$, 
defined by (\ref{5-b}). A set of basis invariants are $x,y, U(x,y)\,=\,
u\,-\,2 \ln z$. The associated reduced equation is the Liouville equation
       \begin{eqalignno}
\partial_{\eta}\partial_{\bar{\eta}}U\,=\,\frac{\kappa}{2}e^{U},
\label{15-b}
\end{eqalignno}
whose general solution is
\begin{eqalignno}
e^{U}\,=\, \frac{f_{\eta}(\eta)\,{\bar f}_{\bar\eta}(\bar \eta)}
{\left[1+\frac{1}{4}f(\eta)\,{\bar f}(\bar \eta)\right]^{2}} \label{15-c}
\end{eqalignno}
\end{subequations}
for $\kappa\,=\,-1$, where $f(\eta)$ denotes an arbitrary holomorphic 
function. The solution of Eq.(1) corresponding to (\ref{15-c}) reads
\begin{subequations} \label{16}
       \begin{eqalignno}
e^{u}\,=\, \frac{z^{2}\,f_{\eta}(\eta)\,{\bar f}_{\bar\eta}(\bar \eta)}
{\left[1+\frac{1}{4}f(\eta)\,{\bar f}(\bar \eta)\right]^{2}} \label{16-a} 
\end{eqalignno}
which produces the Eguchi-Hanson gravitational instanton \cite{Eguchi}
       \begin{eqalignno}
e^{u}\,=\, \frac{4(z^{2}-z^{2}_{0})}
{\left(1+x^{2}+y^{2}\right)^{2}} \label{16-b},
\end{eqalignno}
\end{subequations}
by setting $f(\eta)\,=\,2\eta$, ${\bar f}(\bar\eta)\,=\,2\bar\eta$, 
($z_{0}=\,$ constant).

{}From Eq.(\ref{15-b}) other gravitational configurations of physical 
interest emerge.

By way of example, if $f(\eta)=2 c^{N}\eta^{N}$, $\bar{f}
(\bar\eta)=2\bar{c}^{N}\bar{\eta}^{N}$, we have
\begin{subequations}\label{17}
\begin{eqalignno}
e^{u}\,=\,\frac{4N^{2}z^{2}}{r^{2}}\left[
\left(\frac{r_{0}}{r}\right)^{N}\,+\,\left(\frac{r}{r_{0}}\right)^{N}
\right]^{-2}, \label{17-a}
\end{eqalignno}
where $r_{0}$ and $N$ are two positive constants such that $N\ge 1$. The 
last condition ensures the regularity of the r.h.s. of (\ref{17-a}) at the 
origin and at infinity, where $e^{u}_{r\rightarrow 0} 
\sim r^{2N-2}$ and $e^{u}_{r\rightarrow \infty} 
\sim r^{-2N-2}$, respectively. The expression (\ref{17-a}), which contains 
(\ref{16-b}) as a special case, resembles 
strongly that appearing in the context of the self-dual Chern-Simons model 
developed by Jackiw and Pi \cite{Pi}. In our framework, the quantity 
(\ref{17-a}) can be interpreted as an $N$-gravitational instanton solution 
where the instantons are superimposed at the origin (each with the 
same scale $r_{0}$).

Another nontrivial configuration generated by the symmetry operator 
$Y$ (via the Liouville equation (\ref{15-b})) is determined by (\ref{16-a}) 
for $f(\eta)=2\sum^{N}_{i=1}c_{i}/(\eta-\eta_{i})$, where $c_{i}$ are 
complex constants. It follows that
\begin{eqalignno}
e^{u}\,=\,\frac{4|\sum^{N}_{i=1}\frac{c_{i}}{(\eta-\eta_{i})^{2}}|^{2}\,
z^{2}}{\left[1\,+\,|\sum^{N}_{i=1}\frac{c_{i}}{(\eta-\eta_{i})}|^{2}\,
\right]^{2}},
\label{17-b}\end{eqalignno}
\end{subequations}
which describes separated gravitational instantons depending on $4N$ real 
parameters \cite{Pi}. 

The metrics related to the solutions (\ref{17-a}) and (\ref{17-b}) can be 
straightforwardly calculated by Eq.(2). As one expects, the metric 
associated with (\ref{17-a}) is singularity free and asymptotically flat, 
while the metric 
of (\ref{17-b}) corresponds to that devised by Gibbons and Hawking resorting 
to the multicenter {\sl ansatz} \cite{gibbons}. 

On the other hand, in opposition to what happens 
for the gravitational instantons, the metric 
tensors defining the meron-like solutions present singularities. For 
instance, the metric tensor of the configuration (\ref{13-b}) is 
\begin{eqalignno} 
g=\, &2\,\left\{\frac{\rho^{2}}{x^{2}}\left[(dx)^{2}
+ (dy)^{2}\right]+4(d\rho)^{2}\right\}\nonumber\\
+\,& \frac{1}{2}\rho^{2}\left(-\frac{2\epsilon}{x}
dy + d\tau\right)^{2}, \label{m-3} 
\end{eqalignno}
with $z=\rho^{2}$. The expression (\ref{m-3}) has a singularity at $x=0$. 
This is in accordance with the meron-like character of (\ref{13-b}).

As a final example of remarkable symmetries of Eq.(1),
 we shall deal with the generators 
{\it i)} $V_{0}=T_{0}+Y_{0}$, where $T_{0}$ and $Y_{0}$ are expressed 
by (\ref{10-b}) and (\ref{10-d}), and {\it ii)} $X_{0}=n h\, 
\sqrt{1-{h}^{2}}\,(h\partial_{h}\,+\,l \partial_{l}\,-
\,2\partial_{u}),$ which comes from (\ref{5-a}) by setting $f=e^{nx}
\cos{ny}$, $\hat{f}=e^{nx}
\sin{ny}$ ($h\,=\,e^{nx}, \,l\,=\,\sin{ny}$).

The case $i)$ shows the existence of an unsuspected link between the 
heavenly equation (1) and an equation introduced by Schr{\"o}der in the 
formulation of a combinatorial problem over 100 years ago \cite{Hagedorn}.

Schr{\"o}der's equation appears also in the context of the statistical 
bootstrap model and in renormalization theory \cite{Hagedorn}. A set 
of invariants of $V_{0}$ is $\beta\,={\rm arctan}t-z/\sqrt{2}$, 
$\Lambda(\beta)\,=W(\beta, \frac{z}{\sqrt{2}})-\ln(1+\beta^{2})$. Then, 
Eq.(\ref{9}) yields the ordinary differential equation
\be 
\Lambda_{\beta\beta}\,=\,\kappa (e^{\Lambda})_{\beta\beta},
\lab{18}\ee
which gives the Schr{\"o}der equation
\be
2\Lambda\,-\,e^{\Lambda}\,+\,1\,=\,\beta,
\lab{19}\ee
for $\kappa=1$ and after a suitable choice of the constants of integration. 

The analytical structure of (\ref{19}) was investigated by Hagedorn and 
Rafelsky \cite{Hagedorn} who found an integral representation of 
$\Lambda$ in the principal Riemann sheet.

A solution of Eq.(\ref{19}) in terms of a power series expansion is given 
by \cite{Hagedorn}
\be
\Lambda\,=\,\Lambda_{0}-\,\sum_{n=1}^{\infty}c_{n}(\beta_{0}\,-\,\beta)
^{\frac{1}{n}},
\lab{20}\ee
where $\beta_{0}\,=2\ln{2} -1$ is the point at which $\beta(\Lambda)$ has a 
maximum with value $\Lambda_{0}\,=\,\ln 2$ and the coefficients $c_{n}$ can be 
calculated by the recursion relation
$$
c_{n}\,=\,\frac{1}{2}\left[\frac{n-1}{n+1} c_{n-1}\,-\,
\sum_{k=2}^{n-1}c_{k}c_{n+1-k}
\right]\qquad
  (c_{1}=1).$$
The series converges rapidly in the region $0\le\beta\le\beta_{0}$.
Thus, taking account of (\ref{20}) we get the solution
\be u\,=\,\ln\frac{2}{r^{2}}\,-\,\sum_{n=1}^{\infty}\,c_{n}\left[
\frac{1}{\sqrt{2}}(z-z_{0})-(\theta-\theta_{0}))\right]
^{\frac{1}{n}}\lab{21-a}\ee
to Eq.(1), where $\,\beta\,=\,\theta\,-\,{z}/\sqrt{2}$, $\theta\,=\,{\rm arctan}
\frac{y}{x}$, $r\,=\,\sqrt{x^{2}\,+\,y^{2}}$, and $u=W-2\ln x$.

The case $ii)$ leads to the reduced equation
\be
U_{qq}\,=\,\kappa(e^{U})_{zz},
\lab{22}\ee
where $z,\,p=\frac{1}{n}e^{-nx} \sin{ny}$ and $U(q,z)\,=u+2nx$ are 
invariants. Equation (\ref{22}) is a two-dimensional version of Eq.(1). 
Special solutions to 
Eq.(\ref{22}) are $\,U\,=2\ln\left({z}/{\cos p}\right)\,$, for 
$\kappa\,=\,1$ \cite{APS}, and $U\,=\,2\ln\left({z}/{\cosh p}\right)$ 
for $\kappa\,=\,-1$, so that the functions
\bea
u_{1}\,=&\,2 \ln\frac{z}{\cos p}\,-\,2nx, \;\;
u_{2}\,=&\,2 \ln\frac{z}{\cosh p}\,-\,2nx  \lab{23}
\eea
solve, in correspondence, the original equation (1).

At present, the physical interpretation of the configurations (\ref{21-a}) and 
(\ref{23}) remains to be elucidated, and
deserves further investigation. Here we 
remark only that (\ref{21-a}) constitutes, at the best of our knowledge, 
the only known example of a solution to Eq. (1) in which 
the variable $z$ appears in a 
nontrivial way (usually, the dependence of $e^{u}$ on $z$ looks as a 
polynomial of the second degree).

To conclude this note, we emphasize the role of the symmetry approach in 
the classification of the gravitational excitations allowed by the heavenly 
equation.


\begin{thebibliography}{99}

\bibitem{Saveliev} {See, for example:} M. V. Saveliev, 
\Journal{\em Theor. Math. Fiz.}{92}{457}{1992}, and references 
therein;
M. Przanowski, \Journal{\JMP}{31}{300}{1990}; 
\Journal{\em ibid.}{32}{1004}{1991};
Q-Han Park, \Journal{\PLB}{236}{429}{1990}.
%
\bibitem{Plebanski}J. F. Plebanski, \Journal{\JMP}{16}{2395}{1975}; 
 J. D. Gegenberg and A. Das, 
\Journal{\em Gen. Relativ. Grav.}{16}{817}{1989}; C. Lebrun, 
\Journal{\em J. Diff. Geom.}{34}{223}{1991}.
\bibitem{Eguchi} T. Eguchi, P. B. Gilkey and A. J. Hanson, \Journal{\em Phys. 
Reps.}{66}{213}{1980}.
%
\bibitem{Hawking} S. W. Hawking, in {\em Recent developments in gravity},
edited by  M. Leavy and S. Deser (Plenum, New York, 1979).
%
\bibitem{Olver} P. J. Olver, {\it Applications of Lie groups to differential 
Equations } (Springer, New York, 1986).
%

\bibitem{Hagedorn}R. Hagedorn and J. Rafelski, 
\Journal{\em Comm. Math. Phys.}{83}{563}{1982}, and references therein. 
%
\bibitem{Carmeli2} M. Carmeli, {\em Group theory and general relativity} 
(Mcgraw-Hill, London, 1977) p.30; 
B. G. Wybourne, {\em Classical groups for physicists} 
(J. Wiley and Sons, New York, 1974).
%
\bibitem{Carmeli} M. Carmeli, {\em Classical fields: general relativity and 
field theory} (Mcgraw-Hill, London, 1977).
%
\bibitem{SC}
F. Schwarz, {\em Comput. Phys. Commun.} {\bf 27}, 179 (1982); {\em Computing} 
{\bf 34}, 9 (1985).
%
%
\bibitem{Gatteschi} L. Gatteschi, {\em Funzioni speciali} (UTET, Torino, 
1973); M. Abramowitz  and I. Stegun  (Eds.), {\em Handbook of 
mathematical functions} (Dover, New York, 1975), p.555.
%
\bibitem{Gidas} See, for example, A. Jaffe, {in \em Recent 
developments in gauge theories}, {edited by G. 't Hooft {et. al.}} (Plenum, 
New York, 1979) p.189; A. Jaffe and C. Taubes, {\em Vortices and monopoles} 
(Birkh{\"a}user, Boston, 1980), Chap. 3; F. A. Schaposnik and J. E. 
Solomin, \Journal{\JMP}{20}{2110}{1979}; B. Gidas, 
\Journal{\JMP}{20}{2097}{1979}. 

\bibitem{Cheng} T. Cheng and  L. Li, {\em Gauge field theory of elementary 
particle physics} (Oxford, 1984), p.478. 

\bibitem{Friedlander}F.G. Friedlander,
{\em The wave equation on a curved space-time} (Cambridge University, 
London, 1975), p.57. 
%
\bibitem{Pi} R. Jackiw and S. Y. Pi, \Journal{\PRD}{42}{3500}{1990}. 
%
\bibitem{gibbons}G. W. Gibbons and S. W. Hawking, 
\Journal{\PLB}{78}{430}{1978}.

\bibitem{APS} E. Alfinito, G. Profilo and  G. Soliani, 
University of Lecce - preprint, 1995. 

\end{thebibliography}
\end{document}